\documentclass[%
 reprint,
 superscriptaddress,
 showpacs,preprintnumbers,
 amsmath,amssymb,
 aps,
 prb,
]{revtex4-1}

\usepackage{graphicx}
\usepackage{dcolumn}
\usepackage{bm}

\usepackage{upgreek}
\usepackage{amsmath}

\begin{document}


\title{Gain-assisted critical coupling for enhanced optical absorption in graphene}

\author{Tingting Liu}
\affiliation{School of Physics and Electronics Information, Hubei University of Education, Wuhan 430205, China}

\author{Chaobiao Zhou}
\affiliation{College of Mechanical and Electronic Engineering, Guizhou Minzu University, Guiyang 550025, China}

\author{Shuyuan Xiao}
\email{syxiao@ncu.edu.cn}
\affiliation{Institute for Advanced Study, Nanchang University, Nanchang 330031, China}
\affiliation{Jiangxi Key Laboratory for Microscale Interdisciplinary Study, Nanchang University, Nanchang 330031, China}

\begin{abstract}
	
Enhanced optical absorption in two-dimensional (2D) materials has recently moved into the focus of nanophotonics research. In this work, we present a gain-assisted method to achieve critical coupling and demonstrate the maximum absorption in undoped monolayer graphene in the near-infrared. In a two-port system composed of photonic crystal slab loaded with graphene, the gain medium is introduced to adjust the dissipative rate to match the radiation rate for the critical coupling, which is accessible without changing the original structural geometry. The appropriate tuning of the gain coefficient also enables the critical coupling absorption within a wide wavelength regime for different coupling configurations. This work provides a powerful guide to manipulate light-matter interaction in 2D materials and opens up a new path to design ultra-compact and high-performance 2D material optical devices. 

\end{abstract}

\maketitle


\section{\label{sec1}Introduction}

Optical absorption is a fundamental problem in the scientific community of light-matter interaction because of diverse application scenarios including energy harvesting\cite{Zhou2016, Ma2018}, imaging\cite{Chien2018, Simoncelli2018, Zhou2020}, sensing\cite{Rodrigo2015, Liu2019a, Zhao2020}, photocatalysis\cite{Warren2012, Zhao2018}, and photodetection\cite{Tagliabue2018, Guo2020}. One plausible method of absorption enhancement has been proposed based on the concept of critical coupling\cite{Haus1984, Fan2003}. It can be generalized to the condition that the incident wave energy in a resonant optical structure with some dissipative loss will be fed to the system with maximum efficiency when the leakage rate of energy out of the resonator and the internal resonator loss are equal. This concept has significantly benefited the research field of two dimensional (2D) materials where they are inserted in such a system so that the maximum absorption efficiency is achieved at critical coupling. Such critical coupling mechanism has been first devoted to achieve total absorption in monolayer graphene in the near-infrared in the absence of plasmonic response\cite{Piper2014, Liu2014, Lu2015, Guo2016, Jiang2017, Akhavan2018, Wang2019, Xiao2020, Wang2020a}, and afterwards applied to the research in the whole family of 2D materials, including transitional metal dichalcogenides in the visible\cite{Huang2016, Li2017, Jiang2018, Hong2019, Cao2020, Wang2020}, and black phosphorus in the mid-infrared\cite{Qing2018, Xiao2019, Liu2019, Liu2020, Xu2020}, which provides a simple, universal, and robust strategy for ultra-compact and high-performance optical devices.

In the coupling system for absorption enhancement of 2D materials, the performance is fundamentally determined by how closely the system is approaching the exact critical coupling condition, in which a delicate interaction between the resonance mode and the configurations reaches the balance of radiation and dissipation rates, i.e. $\gamma=\delta$. Such exact critical coupling condition is generally involved with carefully optimizing parameters of the coupled systems by means of time-consuming rigorous numerical simulations. Typically, by tuning the geometrical parameters of the coupling configuration, the radiation rate $\gamma$ can be largely modulated in order to match the dissipative rate $\delta$ which is predominantly determined by the intrinsic material loss. In contrast, an alternative approach consists in the manipulation of the dissipative rate $\delta$ by varying the material permittivity or losses of tunable elements . However, the performance of the precisely designed configurations as well as the radiation and dissipative rates is usually difficult to control once fabricated. In pursuit of practical applications, the availability of active gain medium with optical gain which has been used to compensate the loss due to intrinsic absorption\cite{Zheludev2008, Xiao2010, Yoon2015, Vasic2017, Wang2017, Meng2019, Sanders2020}, makes a compelling case for exploring its efficiency to strictly obtain the critical coupling absorption in 2D materials. 

In this work, we present a generic guidance to enhance optical absorption in 2D materials and achieve the gain-assisted critical coupling for the maximum absorption in undoped monolayer graphene in the near-infrared. In an exemplary coupling system comprising a monolayer graphene on top of a photonic crystal (PhC) slab supporting guided resonance, we show that the optical gain medium can be employed to adjust the dissipative rate $\delta$ on a large scale to match the radiation rate $\gamma$ for the critical coupling, which is precisely accessible without changing the original structural geometry. Even for different geometrical parameters of the system, the appropriate gain coefficient tuning can also achieve the critical coupling within a wide resonance wavelength regime while maintaining the maximum absorption efficiency. This work, formulated in general fashion, should be useful to achieve the critical coupling absorption in 2D material families assisted with active gain. 

\section{\label{sec2}Review of critical coupling mechanism}

The general aspect of critical coupling absorption is revealed by the temporal coupled-mode theory (CMT) which describes the input and output properties of a dissipative resonator. When applying the theory to the two-port system which supports a single resonance, the dynamic equations for resonance amplitude $a$ can be written as\cite{Haus1984, Fan2003},
\begin{eqnarray}
\frac{da}{dt}&=&(i\omega_{0}-\gamma-\delta)a+D^{T}|s_{+}\rangle,\label{eq1} \\
|s_{-}\rangle&=&C|s_{+}\rangle+Da,\label{eq2}
\end{eqnarray}
where $\gamma$ is the total external radiation rate of the resonance system, $\delta$ is the intrinsic dissipative rate of the system, and $\omega_{0}$ is the resonance frequency, $|s_{+}\rangle$, $|s_{-}\rangle$ represent input and output amplitudes from each port, $D$ describes the coupling coefficient between the resonator mode and the ports, $C$ is the scattering matrix for the coupling between the incoming and outgoing waves in the ports through a direct pathway. Constrained by the energy conservation and time-reversal symmetry considerations, the absorption efficiency $A$ in the resonance system can be derived from the above equations as
\begin{eqnarray}
\begin{split}
A
&= \frac{\langle s_{+}|s_{+}\rangle-\langle s_{-}|s_{-}\rangle}{\langle s_{+}|s_{+}\rangle}=\frac{2\delta|a|^{2}}{\langle s_{+}|s_{+}\rangle}      \\
&= \frac{2\delta\gamma}{(\omega-\omega_{0})^{2}+(\gamma+\delta)^{2}}.\label{eq3}
\end{split}
\end{eqnarray} 
We notice that the critical coupling absorption with maximum efficiency can be obtained as $A=0.5$ when the system is driven on resonance with $\omega=\omega_{0}$, and the radiation and dissipative rates are equal, i.e. $\gamma=\delta$. It has also been demonstrated that the total absorption can be further achieved when the two-port system is replaced with the one-port system using a back reflector of a metallic mirror or a multilayer dielectric Bragg mirror\cite{Piper2014}. In addition to the absorption efficiency, the absorption bandwidth which is defined as the full width at half maximum (FWHM) $\Gamma^{\text{FWHM}}$ has also been deduced as
\begin{eqnarray}
\Gamma^{\text{FWHM}}=2(\gamma+\delta).\label{eq4}
\end{eqnarray}
Under the critical coupling condition, the FWHM can be simplified to $4\gamma$ or $4\delta$. Therefore, the manipulation of absorption properties in the coupling system can be achieved by controlling the radiation and dissipative rates.  

\section{\label{sec3}Numerical results and discussions}

\begin{figure}[htbp]
\centering
\includegraphics
[scale=0.40]{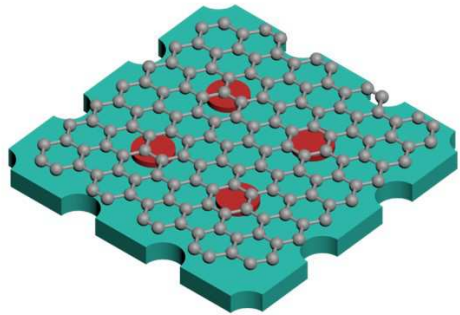}
\caption{\label{fig1} Schematic of the proposed critical coupling system consisting of a PhC slab loaded with undoped monolayer graphene. The gain medium fills the holes.}
\end{figure}

We consider a typical two-port configuration with undoped monolayer graphene on top of a PhC slab composed of a square lattice of cylindrical holes in a sheet of silicon, while a gain medium as the tunable element simply fills the holes, as Fig. \ref{fig1} shows. The geometric parameters of the PhC slab are the period $P=900$ nm, the thickness $t=100$ nm, and the hole radius $r$. The numerical calculations are conducted using the finite-difference time-domain (FDTD) algorithm. In the calculation, the refractive index of silicon is set as $n=3.5$, acting as a lossless material in the wavelength of interest\cite{Palik1998}. The surface conductivity of the undoped monolayer graphene is set as $e^{2}/(4\hbar)$, serving as ultrathin lossy film in visible and near-infrared\cite{Stauber2008, Hanson2008}. The dielectric permittivity of the gain material is described by the Lorentz model\cite{Oughstun2003},
\begin{eqnarray}
\varepsilon=\varepsilon_{0}+\frac{\varepsilon_{\text{gain}}\omega_{\text{gain}}^{2}}{\omega_{\text{gain}}^{2}-2i\omega\delta_{\text{gain}}-\omega^{2}},\label{eq5}
\end{eqnarray}
where $\varepsilon_{0}$ for this dielectric case is set as 1.6, $\omega_{\text{gain}}$ is the gain center frequency, $\delta_{\text{gain}}$ mainly influences the width of the gain spectra. The gain coefficient $g$ is varied by shifting $\varepsilon_{gain}$ with $g=(4\pi/\lambda)\text{Im}(\sqrt{\varepsilon})$, while the sign $\pm$ controls if the material has loss or gain. Here for simplicity, we use the value of $g$ corresponding to the maximum gain value at $\omega_{\text{gain}}$. In the proposed structure, the radiation rate $\gamma$ depends primarily on the geometrical parameters, which implies that $\gamma$ is fixed for a given structure. The dissipative rate $\delta$ is dominantly determined by the loss rate of undoped graphene layer and the growth rate of gain medium. It is desirable to obtain the same $\gamma$ and $\delta$ for critical coupling and then the maximum absorption of undoped graphene by controlling these quantities above.

\begin{figure}[htbp]
\centering
\includegraphics
[scale=0.45]{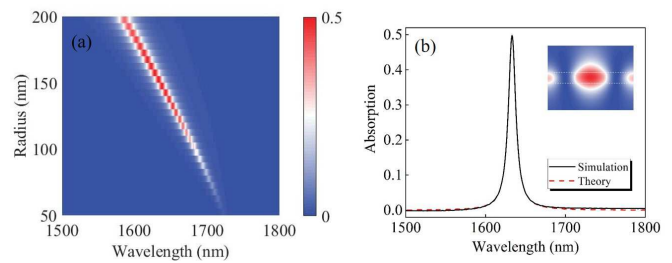}
\caption{\label{fig2} The absorption properties for the proposed structure with $g=0$ cm$^{-1}$. (a) The absorption spectra as a function of incident wavelength and the cylinder hole radius; (b) the simulated and theoretical absorption spectra at critical coupling for the radius $r=150$ nm, and inset is the corresponding electric field distribution of $|E|^2$ on the $y$-$z$ cross-sectional plane through the center of the hole.}
\end{figure}

\begin{figure*}[htbp]
\centering
\includegraphics
[scale=0.90]{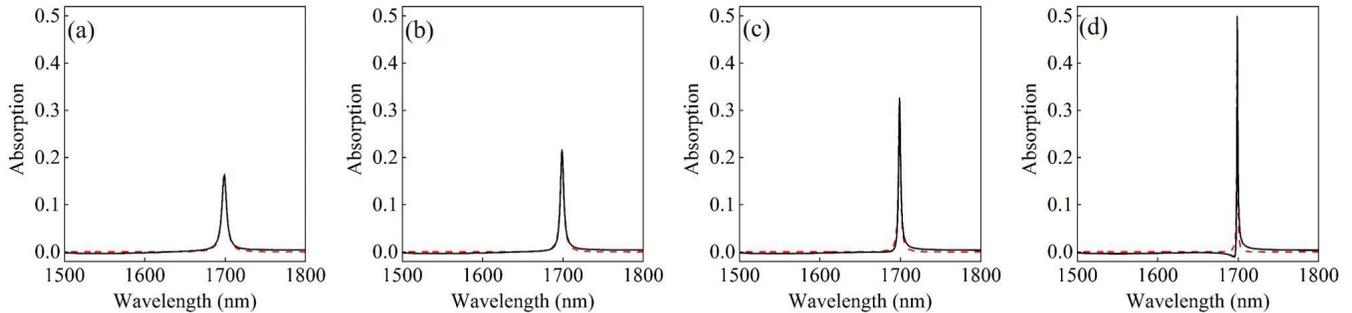}
\caption{\label{fig3}(a)-(d) The simulated and theoretical absorption spectra of the proposed structure with the hole radius $r=80$ nm for different gain coefficients $g=0$ cm$^{-1}$, $-5.15\times10^{3}$ cm$^{-1}$, $-1.02\times10^{4}$ cm$^{-1}$, and $-1.45\times10^{4}$ cm$^{-1}$, respectively.}
\end{figure*}

\begin{figure}[htbp]
\centering
\includegraphics
[scale=0.40]{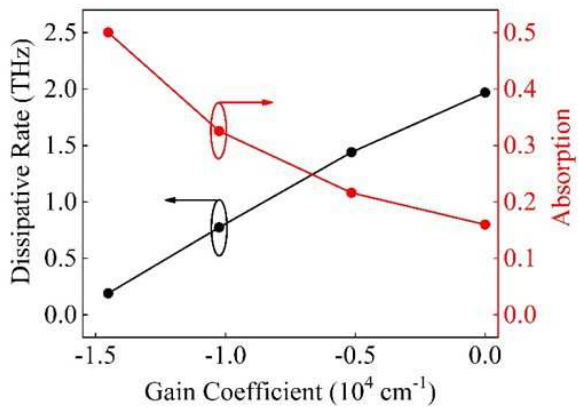}
\caption{\label{fig4} The variations of the dissipative rates of the system and the absorption efficiency of graphene for different gain coefficients.}
\end{figure}

\begin{figure*}
\centering
\includegraphics
[scale=0.90]{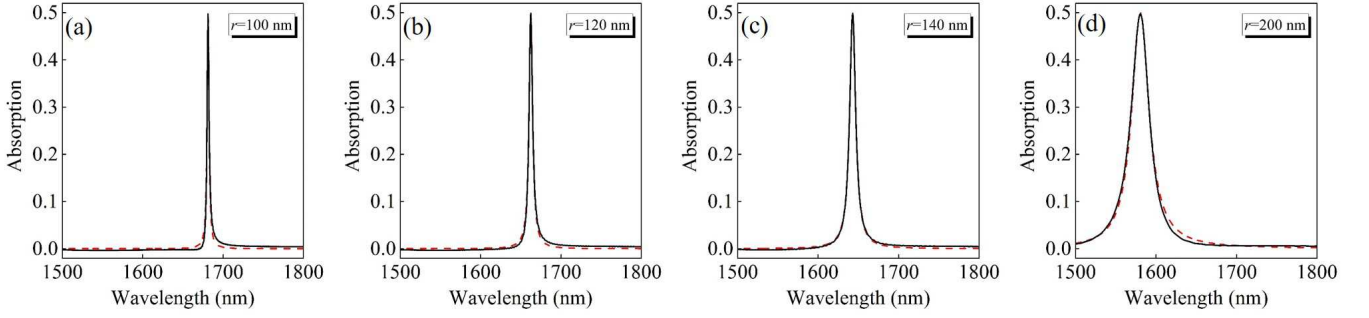}
\caption{\label{fig5} (a)-(d) The simulated and theoretical absorption spectra of the proposed structure with the hole radius $r=100$ nm, 120 nm, 140 nm, and 200 nm for corresponding gain coefficients $g=-9.17\times10^{3}$ cm$^{-1}$, $-5.38\times10^{3}$ cm$^{-1}$, $-1.65\times10^{3}$ cm$^{-1}$, and $7.42\times10^{4}$ cm$^{-1}$, respectively.}
\end{figure*}

In the coupling system, the PhC slab is employed as a resonator where the guided mode couples to the external radiation, forming leaky guided resonance when the in-plane periodicity enables phase-matched coupling\cite{Miroshnichenko2010, Zhou2019}. Accordingly, the Fano resonance with the asymmetric reflection spectra is contributed primarily by the toroidal dipole moment, allowing the confinement of the incident electromagnetic energy within the PhC slab (Fig. S1 of the Supplementary Material). Then the design principle for critical coupling system is fulfilled when the absorbing material undoped graphene is integrated into such a resonant structure. To gain a comprehensive understanding of the critical coupling condition, the absorption spectra of the proposed structure with a gain coefficient $g=0$ cm$^{-1}$ as functions of the hole radius and the incident wavelength are displayed in Fig. \ref{fig2}(a). As a direct consequence of increasing the hole radius, the radiation rate $\gamma$ in the system becomes larger whilst the dissipative rate $\delta$ keeps basically steady for the undoped graphene. From the initial under coupling state with $\gamma<\delta$, the system reaches the critical coupling condition with $\gamma=\delta$ and finally the over coupling state $\gamma>\delta$. Thus, the absorption efficiency firstly increases to the maximum value and then decreases, and the spectral bandwidth gradually increases. Through this process, the critical coupling point, i.e. the maximum absorption of $A=0.5$, can be found at the hole radius of 150 nm. The corresponding simulated absorption spectrum is depicted in Fig. \ref{fig2}(b) while the strongly confined electric field distribution of $|E|^2$ at resonance is shown in the inset. By fitting the simulated result with Eq. (\ref{eq3}), it is possible to exact the radiation and dissipative rates as $\gamma=\delta=1.97$ THz under the critical coupling condition. The maximum absorption can be directly realized by reaching the balance of the $\delta$ and $\gamma$ via altering the geometric parameters in this conventional way. 

In sharp contrast to the approach of above radiation rate tuning, the gain medium which is intuitively opposite to the absorption is introduced to manipulate the dissipative rate of the system and realize the strictly critical coupling absorption in the resonant structure. The tunable ability of the gain medium is demonstrated using the example of the proposed structure with a fixed hole radius $r=80$ nm for different gain coefficients $g$. It has been mentioned above that the system is in the under coupling state with $\gamma<\delta$ for the case of $r=80$ nm and $g=0$ cm$^{-1}$. In order to achieve the critical coupling condition with $\delta=\gamma$, the gain level needs to be increased to compensate the intrinsic loss for altering $\delta$, instead of changing $\gamma$ via conventionally geometric parameter adjustment. For this purpose, the gain coefficient $g$ varies from 0, $-5.15\times10^{3}$ cm$^{-1}$, $-1.02\times10^{4}$ cm$^{-1}$ to $-1.45\times10^{4}$ cm$^{-1}$ as shown in Fig. \ref{fig3}(a)-(d). For the initial states with $g=0$ cm$^{-1}$, the small Fermi surface in undoped graphene restricts the allowed phonon energy in collisions. Then the utilization of gain gives rise to the generation of more photons and slow down of the absorption process, as a result of which the total $\delta$ in the system reduces. Since $\gamma$ is considered as constant due to the unchanged geometry, the approaching $\delta$ to $\gamma$ leads to the evidently increasing absorption efficiency, as shown in Fig. \ref{fig4}. When $\delta$ reaches the same value with $\gamma$, the undoped graphene exhibits the maximum absorption of 0.5 at critical coupling. By continuously increasing gain level, the total $\delta$ would further decreases and become less than $\gamma$ or even become negative, providing new functionality of amplification once passing through the zero point\cite{Li2010, Liu2011}. In particular, the condition of the initiation of lasing is satisfied when the sum $\delta+\gamma$ approaches zero at $g=-2.49\times10^{4}$ cm$^{-1}$ (Fig. S2 of the Supplementary Material). It is interesting that the utilization of gain in the proposed structure not only allows the enhanced graphene absorption via critical coupling, but also enables the new functionality such as amplification or lasing. 

\begin{figure}[htbp]
\centering
\includegraphics
[scale=0.40]{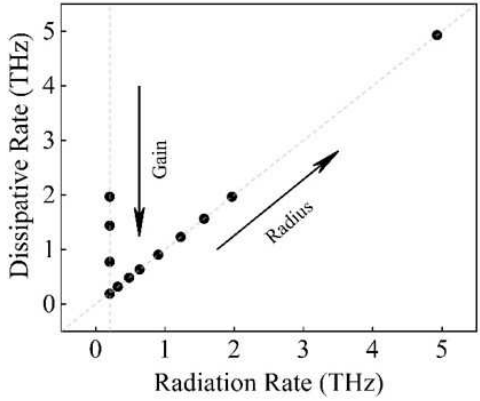}
\caption{\label{fig6} The radiation rate $\gamma$ and dissipative rate $\delta$ of the proposed critical coupling system for different gain coefficients $g$ and hole radii $r$. The arrow directions show the trend of $g$ and $r$, respectively.}
\end{figure}

On the other hand, controlling of the dissipative rate $\delta$ based on the gain-assisted of critical coupling is also investigated for different geometrical parameters of the proposed structure. In Fig. \ref{fig5} where the hole radius $r$ ranges from 100 nm to 200 nm, the maximum absorption can be achieved at different wavelengths by adjusting the gain coefficient. It is observed that the resonance wavelength shows blue shift from 1681.41 nm to 1580.6 nm due to the reducing effective refractive index of the PhC slab as the hole radius $r$ increases. It is also known that the radiation rate $\gamma$ comes closer to the dissipative rate $\delta$ because $\gamma$ increases as the hole radius increases when the gain does not occur for the initial under coupling case. Hence the required gain level is becoming lower to reach the balance between $\delta$ and $\gamma$. Meanwhile, the absorption bandwidth in the graphene also shows increase due to the simultaneous shift of $\delta$ and $\gamma$, i.e., $\Gamma^{\text{FWHM}}$ increases from 2.88 nm to 26.15 nm for the increasing radius, which has been well predicted by Eq. (\ref{eq4}). Moreover, the continuous increase of the radius results in the over coupling state with larger $\gamma$ than $\delta$. In this case, instead of the gain with negative $g$ value, the lossy material with positive $g$ is adopted to maintain the equal values of the rates, for example, $g=7.42\times10^{4}$ cm$^{-1}$ is used for the critical coupling at radius of 200 nm. In practice, the absorption characteristics of the proposed critical coupling system will be fixed once the structure is fabricated. The feasibility of gain at different wavelengths offers great possibility to enhance and maximize graphene absorption via gain-assisted critical coupling within a wide wavelength range.

Fig. \ref{fig6} summarizes the results from those in Fig. \ref{fig3} and Fig. \ref{fig5}, and provides the variations of radiation and dissipative rates with different gain coefficients and radii. For the given structure in Fig. \ref{fig3}, the system is initially in the under coupling state with the gain coefficient $g=0$ cm$^{-1}$ due to large $\delta$ and small $\gamma$. As the gain medium take effects with increasing $g$, the total $\delta$ of the coupling system gradually decreases and then reaches the same value with $\gamma$ at a certain value, which is about $10\%$ of the original $\delta$ without gain. At this point, the system realizes the critical coupling and the undoped graphene obtain maximum absorption efficiency. Hence for the given under coupling system with $\delta>\gamma$, the absorption can be increased as $\delta$ approaches $\gamma$ by means of increasing optical gain. On the other hand, the radiation and dissipative rates at critical coupling for different radii are calculated from the results in Fig. \ref{fig5} and depicted in Fig. \ref{fig6} as well. Since $\gamma$ becomes larger as the hole radius increases, $\delta$ needs to be modulated to match $\gamma$ for critical coupling. When the radius is 150 nm, this is the case of the critical coupling state with gain coefficient $g=0$ cm$^{-1}$ and the $\delta$ arising from graphene loss is the same with $\gamma$. For the under coupling cases with radius less than 150 nm, the gain medium comes into force with negative coefficient $g$ to reduce the original $\delta$ to reach the balance with $\gamma$. In contrast, once the radius is larger than 150 nm, the gain medium should be replaced with the lossy material with positive $g$ to maintain the equal value with larger $\gamma$. During these modulations, the required gain is within a moderate value at the order of $10^{4}$ cm$^{-1}$. The present method shows the gain has remarkable capability to maximize absorption of graphene in resonant structures by adjustment of $g$ under either the over coupling or under coupling cases. 

\section{\label{sec4}Conclusions}

In conclusions, we present the gain-assisted critical coupling mechanism for the control of optical absorption in 2D materials. This general theory is demonstrated in a two-port coupling configuration with PhC slab loaded with undoped monolayer graphene. Assisted by the gain with a moderate gain coefficient at the order of $10^{4}$ cm$^{-1}$ for the adjustment of the dissipative rate, the absorption efficiency of undoped graphene can be greatly enhanced and maximized in the near-infrared, accompanied with the manipulation of the bandwidth. The appropriate tuning of the gain coefficient also realizes the critical coupling absorption within a wide wavelength regime for different structural geometries. The proposed coupling system can be extended to incorporate other 2D materials in other frequency regimes by re-scaling the PhC slab and choosing appropriate optical gain medium. With this consideration, this work provides a novel and universal guidance for designing high-efficiency 2D material-based optical devices. 

\begin{acknowledgments}	
	
This work is supported by the National Natural Science Foundation of China (Grants No. 61775064, No. 11847132, No. 11947065, No. 61901164, and No. 12004084), the Natural Science Foundation of Jiangxi Province (Grant No. 20202BAB211007), the Interdisciplinary Innovation Fund of Nanchang University (Grant No. 2019-9166-27060003), the Natural Science Research Project of Guizhou Minzu University (Grant No. GZMU[2019]YB22), and the China Scholarship Council (Grant No. 202008420045). The authors would also like to thank Dr. S. Li for her guidance on the effective multipole expansion and Dr. X. Jiang for beneficial discussions on the critical coupling mechanism.

\end{acknowledgments}


\begin{thebibliography}{49}%
	\makeatletter
	\providecommand \@ifxundefined [1]{%
		\@ifx{#1\undefined}
	}%
	\providecommand \@ifnum [1]{%
		\ifnum #1\expandafter \@firstoftwo
		\else \expandafter \@secondoftwo
		\fi
	}%
	\providecommand \@ifx [1]{%
		\ifx #1\expandafter \@firstoftwo
		\else \expandafter \@secondoftwo
		\fi
	}%
	\providecommand \natexlab [1]{#1}%
	\providecommand \enquote  [1]{``#1''}%
	\providecommand \bibnamefont  [1]{#1}%
	\providecommand \bibfnamefont [1]{#1}%
	\providecommand \citenamefont [1]{#1}%
	\providecommand \href@noop [0]{\@secondoftwo}%
	\providecommand \href [0]{\begingroup \@sanitize@url \@href}%
	\providecommand \@href[1]{\@@startlink{#1}\@@href}%
	\providecommand \@@href[1]{\endgroup#1\@@endlink}%
	\providecommand \@sanitize@url [0]{\catcode `\\12\catcode `\$12\catcode
		`\&12\catcode `\#12\catcode `\^12\catcode `\_12\catcode `\%12\relax}%
	\providecommand \@@startlink[1]{}%
	\providecommand \@@endlink[0]{}%
	\providecommand \url  [0]{\begingroup\@sanitize@url \@url }%
	\providecommand \@url [1]{\endgroup\@href {#1}{\urlprefix }}%
	\providecommand \urlprefix  [0]{URL }%
	\providecommand \Eprint [0]{\href }%
	\providecommand \doibase [0]{http://dx.doi.org/}%
	\providecommand \selectlanguage [0]{\@gobble}%
	\providecommand \bibinfo  [0]{\@secondoftwo}%
	\providecommand \bibfield  [0]{\@secondoftwo}%
	\providecommand \translation [1]{[#1]}%
	\providecommand \BibitemOpen [0]{}%
	\providecommand \bibitemStop [0]{}%
	\providecommand \bibitemNoStop [0]{.\EOS\space}%
	\providecommand \EOS [0]{\spacefactor3000\relax}%
	\providecommand \BibitemShut  [1]{\csname bibitem#1\endcsname}%
	\let\auto@bib@innerbib\@empty
	\bibitem [{\citenamefont {Zhou}\ \emph {et~al.}(2016)\citenamefont {Zhou},
		\citenamefont {Tan}, \citenamefont {Wang}, \citenamefont {Xu}, \citenamefont
		{Yuan}, \citenamefont {Cai}, \citenamefont {Zhu},\ and\ \citenamefont
		{Zhu}}]{Zhou2016}%
	\BibitemOpen
	\bibfield  {author} {\bibinfo {author} {\bibfnamefont {L.}~\bibnamefont
			{Zhou}}, \bibinfo {author} {\bibfnamefont {Y.}~\bibnamefont {Tan}}, \bibinfo
		{author} {\bibfnamefont {J.}~\bibnamefont {Wang}}, \bibinfo {author}
		{\bibfnamefont {W.}~\bibnamefont {Xu}}, \bibinfo {author} {\bibfnamefont
			{Y.}~\bibnamefont {Yuan}}, \bibinfo {author} {\bibfnamefont {W.}~\bibnamefont
			{Cai}}, \bibinfo {author} {\bibfnamefont {S.}~\bibnamefont {Zhu}}, \ and\
		\bibinfo {author} {\bibfnamefont {J.}~\bibnamefont {Zhu}},\ }\href {\doibase
		10.1038/nphoton.2016.75} {\bibfield  {journal} {\bibinfo  {journal} {Nat.
				Photonics}\ }\textbf {\bibinfo {volume} {10}},\ \bibinfo {pages} {393}
		(\bibinfo {year} {2016})}\BibitemShut {NoStop}%
	\bibitem [{\citenamefont {Ma}\ \emph {et~al.}(2018)\citenamefont {Ma},
		\citenamefont {Yan}, \citenamefont {Huang}, \citenamefont {Wang},\ and\
		\citenamefont {Yang}}]{Ma2018}%
	\BibitemOpen
	\bibfield  {author} {\bibinfo {author} {\bibfnamefont {C.}~\bibnamefont
			{Ma}}, \bibinfo {author} {\bibfnamefont {J.}~\bibnamefont {Yan}}, \bibinfo
		{author} {\bibfnamefont {Y.}~\bibnamefont {Huang}}, \bibinfo {author}
		{\bibfnamefont {C.}~\bibnamefont {Wang}}, \ and\ \bibinfo {author}
		{\bibfnamefont {G.}~\bibnamefont {Yang}},\ }\href {\doibase
		10.1126/sciadv.aas9894} {\bibfield  {journal} {\bibinfo  {journal} {Sci.
				Adv.}\ }\textbf {\bibinfo {volume} {4}},\ \bibinfo {pages} {eaas9894}
		(\bibinfo {year} {2018})}\BibitemShut {NoStop}%
	\bibitem [{\citenamefont {Chien}\ \emph {et~al.}(2018)\citenamefont {Chien},
		\citenamefont {Brameshuber}, \citenamefont {Rossboth}, \citenamefont
		{Schutz},\ and\ \citenamefont {Schmid}}]{Chien2018}%
	\BibitemOpen
	\bibfield  {author} {\bibinfo {author} {\bibfnamefont {M.-H.}\ \bibnamefont
			{Chien}}, \bibinfo {author} {\bibfnamefont {M.}~\bibnamefont {Brameshuber}},
		\bibinfo {author} {\bibfnamefont {B.~K.}\ \bibnamefont {Rossboth}}, \bibinfo
		{author} {\bibfnamefont {G.~J.}\ \bibnamefont {Schutz}}, \ and\ \bibinfo
		{author} {\bibfnamefont {S.}~\bibnamefont {Schmid}},\ }\href {\doibase
		10.1073/pnas.1804174115} {\bibfield  {journal} {\bibinfo  {journal} {PNAS}\
		}\textbf {\bibinfo {volume} {115}},\ \bibinfo {pages} {11150} (\bibinfo
		{year} {2018})}\BibitemShut {NoStop}%
	\bibitem [{\citenamefont {Simoncelli}\ \emph {et~al.}(2018)\citenamefont
		{Simoncelli}, \citenamefont {Li}, \citenamefont {Cort{\'{e}}s},\ and\
		\citenamefont {Maier}}]{Simoncelli2018}%
	\BibitemOpen
	\bibfield  {author} {\bibinfo {author} {\bibfnamefont {S.}~\bibnamefont
			{Simoncelli}}, \bibinfo {author} {\bibfnamefont {Y.}~\bibnamefont {Li}},
		\bibinfo {author} {\bibfnamefont {E.}~\bibnamefont {Cort{\'{e}}s}}, \ and\
		\bibinfo {author} {\bibfnamefont {S.~A.}\ \bibnamefont {Maier}},\ }\href
	{\doibase 10.1021/acs.nanolett.8b00302} {\bibfield  {journal} {\bibinfo
			{journal} {Nano Lett.}\ }\textbf {\bibinfo {volume} {18}},\ \bibinfo {pages}
		{3400} (\bibinfo {year} {2018})}\BibitemShut {NoStop}%
	\bibitem [{\citenamefont {Zhou}\ \emph {et~al.}(2020)\citenamefont {Zhou},
		\citenamefont {Qu}, \citenamefont {Xiao},\ and\ \citenamefont
		{Fan}}]{Zhou2020}%
	\BibitemOpen
	\bibfield  {author} {\bibinfo {author} {\bibfnamefont {C.}~\bibnamefont
			{Zhou}}, \bibinfo {author} {\bibfnamefont {X.}~\bibnamefont {Qu}}, \bibinfo
		{author} {\bibfnamefont {S.}~\bibnamefont {Xiao}}, \ and\ \bibinfo {author}
		{\bibfnamefont {M.}~\bibnamefont {Fan}},\ }\href {\doibase
		10.1103/physrevapplied.14.044009} {\bibfield  {journal} {\bibinfo  {journal}
			{Phys. Rev. Appl}\ }\textbf {\bibinfo {volume} {14}},\ \bibinfo {pages}
		{044009} (\bibinfo {year} {2020})}\BibitemShut {NoStop}%
	\bibitem [{\citenamefont {Rodrigo}\ \emph {et~al.}(2015)\citenamefont
		{Rodrigo}, \citenamefont {Limaj}, \citenamefont {Janner}, \citenamefont
		{Etezadi}, \citenamefont {de~Abajo}, \citenamefont {Pruneri},\ and\
		\citenamefont {Altug}}]{Rodrigo2015}%
	\BibitemOpen
	\bibfield  {author} {\bibinfo {author} {\bibfnamefont {D.}~\bibnamefont
			{Rodrigo}}, \bibinfo {author} {\bibfnamefont {O.}~\bibnamefont {Limaj}},
		\bibinfo {author} {\bibfnamefont {D.}~\bibnamefont {Janner}}, \bibinfo
		{author} {\bibfnamefont {D.}~\bibnamefont {Etezadi}}, \bibinfo {author}
		{\bibfnamefont {F.~J.~G.}\ \bibnamefont {de~Abajo}}, \bibinfo {author}
		{\bibfnamefont {V.}~\bibnamefont {Pruneri}}, \ and\ \bibinfo {author}
		{\bibfnamefont {H.}~\bibnamefont {Altug}},\ }\href {\doibase
		10.1126/science.aab2051} {\bibfield  {journal} {\bibinfo  {journal}
			{Science}\ }\textbf {\bibinfo {volume} {349}},\ \bibinfo {pages} {165}
		(\bibinfo {year} {2015})}\BibitemShut {NoStop}%
	\bibitem [{\citenamefont {Liu}\ \emph {et~al.}(2019{\natexlab{a}})\citenamefont
		{Liu}, \citenamefont {Liu}, \citenamefont {Tang}, \citenamefont {Liu},
		\citenamefont {Fu},\ and\ \citenamefont {Liu}}]{Liu2019a}%
	\BibitemOpen
	\bibfield  {author} {\bibinfo {author} {\bibfnamefont {G.}~\bibnamefont
			{Liu}}, \bibinfo {author} {\bibfnamefont {Y.}~\bibnamefont {Liu}}, \bibinfo
		{author} {\bibfnamefont {L.}~\bibnamefont {Tang}}, \bibinfo {author}
		{\bibfnamefont {X.}~\bibnamefont {Liu}}, \bibinfo {author} {\bibfnamefont
			{G.}~\bibnamefont {Fu}}, \ and\ \bibinfo {author} {\bibfnamefont
			{Z.}~\bibnamefont {Liu}},\ }\href {\doibase 10.1515/nanoph-2019-0078}
	{\bibfield  {journal} {\bibinfo  {journal} {Nanophotonics}\ }\textbf
		{\bibinfo {volume} {8}},\ \bibinfo {pages} {1095} (\bibinfo {year}
		{2019}{\natexlab{a}})}\BibitemShut {NoStop}%
	\bibitem [{\citenamefont {Zhao}\ \emph {et~al.}(2020)\citenamefont {Zhao},
		\citenamefont {Zhang}, \citenamefont {Yang}, \citenamefont {Li},
		\citenamefont {Feng}, \citenamefont {Quan}, \citenamefont {Yang},\ and\
		\citenamefont {Xiao}}]{Zhao2020}%
	\BibitemOpen
	\bibfield  {author} {\bibinfo {author} {\bibfnamefont {W.}~\bibnamefont
			{Zhao}}, \bibinfo {author} {\bibfnamefont {Y.}~\bibnamefont {Zhang}},
		\bibinfo {author} {\bibfnamefont {J.}~\bibnamefont {Yang}}, \bibinfo {author}
		{\bibfnamefont {J.}~\bibnamefont {Li}}, \bibinfo {author} {\bibfnamefont
			{Y.}~\bibnamefont {Feng}}, \bibinfo {author} {\bibfnamefont {M.}~\bibnamefont
			{Quan}}, \bibinfo {author} {\bibfnamefont {Z.}~\bibnamefont {Yang}}, \ and\
		\bibinfo {author} {\bibfnamefont {S.}~\bibnamefont {Xiao}},\ }\href {\doibase
		10.1039/d0nr02972f} {\bibfield  {journal} {\bibinfo  {journal} {Nanoscale}\
		}\textbf {\bibinfo {volume} {12}},\ \bibinfo {pages} {18056} (\bibinfo {year}
		{2020})}\BibitemShut {NoStop}%
	\bibitem [{\citenamefont {Warren}\ and\ \citenamefont
		{Thimsen}(2012)}]{Warren2012}%
	\BibitemOpen
	\bibfield  {author} {\bibinfo {author} {\bibfnamefont {S.~C.}\ \bibnamefont
			{Warren}}\ and\ \bibinfo {author} {\bibfnamefont {E.}~\bibnamefont
			{Thimsen}},\ }\href {\doibase 10.1039/c1ee02875h} {\bibfield  {journal}
		{\bibinfo  {journal} {Energy Environ. Sci.}\ }\textbf {\bibinfo {volume}
			{5}},\ \bibinfo {pages} {5133} (\bibinfo {year} {2012})}\BibitemShut
	{NoStop}%
	\bibitem [{\citenamefont {Zhao}\ \emph {et~al.}(2018)\citenamefont {Zhao},
		\citenamefont {Xiao}, \citenamefont {Zhang}, \citenamefont {Pan},
		\citenamefont {Wen}, \citenamefont {Qian}, \citenamefont {Wang},
		\citenamefont {Cao}, \citenamefont {He}, \citenamefont {Quan},\ and\
		\citenamefont {Yang}}]{Zhao2018}%
	\BibitemOpen
	\bibfield  {author} {\bibinfo {author} {\bibfnamefont {W.}~\bibnamefont
			{Zhao}}, \bibinfo {author} {\bibfnamefont {S.}~\bibnamefont {Xiao}}, \bibinfo
		{author} {\bibfnamefont {Y.}~\bibnamefont {Zhang}}, \bibinfo {author}
		{\bibfnamefont {D.}~\bibnamefont {Pan}}, \bibinfo {author} {\bibfnamefont
			{J.}~\bibnamefont {Wen}}, \bibinfo {author} {\bibfnamefont {X.}~\bibnamefont
			{Qian}}, \bibinfo {author} {\bibfnamefont {D.}~\bibnamefont {Wang}}, \bibinfo
		{author} {\bibfnamefont {H.}~\bibnamefont {Cao}}, \bibinfo {author}
		{\bibfnamefont {W.}~\bibnamefont {He}}, \bibinfo {author} {\bibfnamefont
			{M.}~\bibnamefont {Quan}}, \ and\ \bibinfo {author} {\bibfnamefont
			{Z.}~\bibnamefont {Yang}},\ }\href {\doibase 10.1039/c8nr02669f} {\bibfield
		{journal} {\bibinfo  {journal} {Nanoscale}\ }\textbf {\bibinfo {volume}
			{10}},\ \bibinfo {pages} {14220} (\bibinfo {year} {2018})}\BibitemShut
	{NoStop}%
	\bibitem [{\citenamefont {Tagliabue}\ \emph {et~al.}(2018)\citenamefont
		{Tagliabue}, \citenamefont {Jermyn}, \citenamefont {Sundararaman},
		\citenamefont {Welch}, \citenamefont {DuChene}, \citenamefont {Pala},
		\citenamefont {Davoyan}, \citenamefont {Narang},\ and\ \citenamefont
		{Atwater}}]{Tagliabue2018}%
	\BibitemOpen
	\bibfield  {author} {\bibinfo {author} {\bibfnamefont {G.}~\bibnamefont
			{Tagliabue}}, \bibinfo {author} {\bibfnamefont {A.~S.}\ \bibnamefont
			{Jermyn}}, \bibinfo {author} {\bibfnamefont {R.}~\bibnamefont
			{Sundararaman}}, \bibinfo {author} {\bibfnamefont {A.~J.}\ \bibnamefont
			{Welch}}, \bibinfo {author} {\bibfnamefont {J.~S.}\ \bibnamefont {DuChene}},
		\bibinfo {author} {\bibfnamefont {R.}~\bibnamefont {Pala}}, \bibinfo {author}
		{\bibfnamefont {A.~R.}\ \bibnamefont {Davoyan}}, \bibinfo {author}
		{\bibfnamefont {P.}~\bibnamefont {Narang}}, \ and\ \bibinfo {author}
		{\bibfnamefont {H.~A.}\ \bibnamefont {Atwater}},\ }\href {\doibase
		10.1038/s41467-018-05968-x} {\bibfield  {journal} {\bibinfo  {journal} {Nat.
				Commun.}\ }\textbf {\bibinfo {volume} {9}},\ \bibinfo {pages} {465205}
		(\bibinfo {year} {2018})}\BibitemShut {NoStop}%
	\bibitem [{\citenamefont {Guo}\ \emph {et~al.}(2020)\citenamefont {Guo},
		\citenamefont {Li}, \citenamefont {Liu}, \citenamefont {Yin}, \citenamefont
		{Wang}, \citenamefont {Ni}, \citenamefont {Fu}, \citenamefont {Yu},
		\citenamefont {Xu}, \citenamefont {Shi}, \citenamefont {Ma}, \citenamefont
		{Gao}, \citenamefont {Tong},\ and\ \citenamefont {Dai}}]{Guo2020}%
	\BibitemOpen
	\bibfield  {author} {\bibinfo {author} {\bibfnamefont {J.}~\bibnamefont
			{Guo}}, \bibinfo {author} {\bibfnamefont {J.}~\bibnamefont {Li}}, \bibinfo
		{author} {\bibfnamefont {C.}~\bibnamefont {Liu}}, \bibinfo {author}
		{\bibfnamefont {Y.}~\bibnamefont {Yin}}, \bibinfo {author} {\bibfnamefont
			{W.}~\bibnamefont {Wang}}, \bibinfo {author} {\bibfnamefont {Z.}~\bibnamefont
			{Ni}}, \bibinfo {author} {\bibfnamefont {Z.}~\bibnamefont {Fu}}, \bibinfo
		{author} {\bibfnamefont {H.}~\bibnamefont {Yu}}, \bibinfo {author}
		{\bibfnamefont {Y.}~\bibnamefont {Xu}}, \bibinfo {author} {\bibfnamefont
			{Y.}~\bibnamefont {Shi}}, \bibinfo {author} {\bibfnamefont {Y.}~\bibnamefont
			{Ma}}, \bibinfo {author} {\bibfnamefont {S.}~\bibnamefont {Gao}}, \bibinfo
		{author} {\bibfnamefont {L.}~\bibnamefont {Tong}}, \ and\ \bibinfo {author}
		{\bibfnamefont {D.}~\bibnamefont {Dai}},\ }\href {\doibase
		10.1038/s41377-020-0263-6} {\bibfield  {journal} {\bibinfo  {journal} {Light
				Sci. {\&} Appl.}\ }\textbf {\bibinfo {volume} {9}},\ \bibinfo {pages} {29}
		(\bibinfo {year} {2020})}\BibitemShut {NoStop}%
	\bibitem [{\citenamefont {Haus}(1984)}]{Haus1984}%
	\BibitemOpen
	\bibfield  {author} {\bibinfo {author} {\bibfnamefont {H.~A.}\ \bibnamefont
			{Haus}},\ }\href@noop {} {\emph {\bibinfo {title} {Waves and Fields in
				Optoelectronics}}}\ (\bibinfo  {publisher} {Prentice-Hall},\ \bibinfo {year}
	{1984})\BibitemShut {NoStop}%
	\bibitem [{\citenamefont {Fan}\ \emph {et~al.}(2003)\citenamefont {Fan},
		\citenamefont {Suh},\ and\ \citenamefont {Joannopoulos}}]{Fan2003}%
	\BibitemOpen
	\bibfield  {author} {\bibinfo {author} {\bibfnamefont {S.}~\bibnamefont
			{Fan}}, \bibinfo {author} {\bibfnamefont {W.}~\bibnamefont {Suh}}, \ and\
		\bibinfo {author} {\bibfnamefont {J.~D.}\ \bibnamefont {Joannopoulos}},\
	}\href {\doibase 10.1364/josaa.20.000569} {\bibfield  {journal} {\bibinfo
			{journal} {J. Opt. Soc. Am. A}\ }\textbf {\bibinfo {volume} {20}},\ \bibinfo
		{pages} {569} (\bibinfo {year} {2003})}\BibitemShut {NoStop}%
	\bibitem [{\citenamefont {Piper}\ and\ \citenamefont {Fan}(2014)}]{Piper2014}%
	\BibitemOpen
	\bibfield  {author} {\bibinfo {author} {\bibfnamefont {J.~R.}\ \bibnamefont
			{Piper}}\ and\ \bibinfo {author} {\bibfnamefont {S.}~\bibnamefont {Fan}},\
	}\href {\doibase 10.1021/ph400090p} {\bibfield  {journal} {\bibinfo
			{journal} {{ACS} Photonics}\ }\textbf {\bibinfo {volume} {1}},\ \bibinfo
		{pages} {347} (\bibinfo {year} {2014})}\BibitemShut {NoStop}%
	\bibitem [{\citenamefont {Liu}\ \emph {et~al.}(2014)\citenamefont {Liu},
		\citenamefont {Chadha}, \citenamefont {Zhao}, \citenamefont {Piper},
		\citenamefont {Jia}, \citenamefont {Shuai}, \citenamefont {Menon},
		\citenamefont {Yang}, \citenamefont {Ma}, \citenamefont {Fan}, \citenamefont
		{Xia},\ and\ \citenamefont {Zhou}}]{Liu2014}%
	\BibitemOpen
	\bibfield  {author} {\bibinfo {author} {\bibfnamefont {Y.}~\bibnamefont
			{Liu}}, \bibinfo {author} {\bibfnamefont {A.}~\bibnamefont {Chadha}},
		\bibinfo {author} {\bibfnamefont {D.}~\bibnamefont {Zhao}}, \bibinfo {author}
		{\bibfnamefont {J.~R.}\ \bibnamefont {Piper}}, \bibinfo {author}
		{\bibfnamefont {Y.}~\bibnamefont {Jia}}, \bibinfo {author} {\bibfnamefont
			{Y.}~\bibnamefont {Shuai}}, \bibinfo {author} {\bibfnamefont
			{L.}~\bibnamefont {Menon}}, \bibinfo {author} {\bibfnamefont
			{H.}~\bibnamefont {Yang}}, \bibinfo {author} {\bibfnamefont {Z.}~\bibnamefont
			{Ma}}, \bibinfo {author} {\bibfnamefont {S.}~\bibnamefont {Fan}}, \bibinfo
		{author} {\bibfnamefont {F.}~\bibnamefont {Xia}}, \ and\ \bibinfo {author}
		{\bibfnamefont {W.}~\bibnamefont {Zhou}},\ }\href {\doibase
		10.1063/1.4901181} {\bibfield  {journal} {\bibinfo  {journal} {Appl. Phys.
				Lett.}\ }\textbf {\bibinfo {volume} {105}},\ \bibinfo {pages} {181105}
		(\bibinfo {year} {2014})}\BibitemShut {NoStop}%
	\bibitem [{\citenamefont {Lu}\ \emph {et~al.}(2015)\citenamefont {Lu},
		\citenamefont {Cumming},\ and\ \citenamefont {Gu}}]{Lu2015}%
	\BibitemOpen
	\bibfield  {author} {\bibinfo {author} {\bibfnamefont {H.}~\bibnamefont
			{Lu}}, \bibinfo {author} {\bibfnamefont {B.~P.}\ \bibnamefont {Cumming}}, \
		and\ \bibinfo {author} {\bibfnamefont {M.}~\bibnamefont {Gu}},\ }\href
	{\doibase 10.1364/ol.40.003647} {\bibfield  {journal} {\bibinfo  {journal}
			{Opt. Lett.}\ }\textbf {\bibinfo {volume} {40}},\ \bibinfo {pages} {3647}
		(\bibinfo {year} {2015})}\BibitemShut {NoStop}%
	\bibitem [{\citenamefont {Guo}\ \emph {et~al.}(2016)\citenamefont {Guo},
		\citenamefont {Zhu}, \citenamefont {Yuan}, \citenamefont {Ye}, \citenamefont
		{Liu}, \citenamefont {Zhang}, \citenamefont {Xu},\ and\ \citenamefont
		{Qin}}]{Guo2016}%
	\BibitemOpen
	\bibfield  {author} {\bibinfo {author} {\bibfnamefont {C.-C.}\ \bibnamefont
			{Guo}}, \bibinfo {author} {\bibfnamefont {Z.-H.}\ \bibnamefont {Zhu}},
		\bibinfo {author} {\bibfnamefont {X.-D.}\ \bibnamefont {Yuan}}, \bibinfo
		{author} {\bibfnamefont {W.-M.}\ \bibnamefont {Ye}}, \bibinfo {author}
		{\bibfnamefont {K.}~\bibnamefont {Liu}}, \bibinfo {author} {\bibfnamefont
			{J.-F.}\ \bibnamefont {Zhang}}, \bibinfo {author} {\bibfnamefont
			{W.}~\bibnamefont {Xu}}, \ and\ \bibinfo {author} {\bibfnamefont {S.-Q.}\
			\bibnamefont {Qin}},\ }\href {\doibase 10.1002/adom.201600481} {\bibfield
		{journal} {\bibinfo  {journal} {Adv. Opt. Mater.}\ }\textbf {\bibinfo
			{volume} {4}},\ \bibinfo {pages} {1955} (\bibinfo {year} {2016})}\BibitemShut
	{NoStop}%
	\bibitem [{\citenamefont {Jiang}\ \emph {et~al.}(2017)\citenamefont {Jiang},
		\citenamefont {Wang}, \citenamefont {Xiao}, \citenamefont {Yan},\ and\
		\citenamefont {Cheng}}]{Jiang2017}%
	\BibitemOpen
	\bibfield  {author} {\bibinfo {author} {\bibfnamefont {X.}~\bibnamefont
			{Jiang}}, \bibinfo {author} {\bibfnamefont {T.}~\bibnamefont {Wang}},
		\bibinfo {author} {\bibfnamefont {S.}~\bibnamefont {Xiao}}, \bibinfo {author}
		{\bibfnamefont {X.}~\bibnamefont {Yan}}, \ and\ \bibinfo {author}
		{\bibfnamefont {L.}~\bibnamefont {Cheng}},\ }\href {\doibase
		10.1364/oe.25.027028} {\bibfield  {journal} {\bibinfo  {journal} {Opt.
				Express}\ }\textbf {\bibinfo {volume} {25}},\ \bibinfo {pages} {27028}
		(\bibinfo {year} {2017})}\BibitemShut {NoStop}%
	\bibitem [{\citenamefont {Akhavan}\ \emph {et~al.}(2018)\citenamefont
		{Akhavan}, \citenamefont {Abdolhosseini}, \citenamefont {Ghafoorifard},\ and\
		\citenamefont {Habibiyan}}]{Akhavan2018}%
	\BibitemOpen
	\bibfield  {author} {\bibinfo {author} {\bibfnamefont {A.}~\bibnamefont
			{Akhavan}}, \bibinfo {author} {\bibfnamefont {S.}~\bibnamefont
			{Abdolhosseini}}, \bibinfo {author} {\bibfnamefont {H.}~\bibnamefont
			{Ghafoorifard}}, \ and\ \bibinfo {author} {\bibfnamefont {H.}~\bibnamefont
			{Habibiyan}},\ }\href {\doibase 10.1109/jlt.2018.2876374} {\bibfield
		{journal} {\bibinfo  {journal} {J. Lightw. Technol.}\ }\textbf {\bibinfo
			{volume} {36}},\ \bibinfo {pages} {5593} (\bibinfo {year}
		{2018})}\BibitemShut {NoStop}%
	\bibitem [{\citenamefont {Wang}\ \emph {et~al.}(2019)\citenamefont {Wang},
		\citenamefont {Chen}, \citenamefont {Zhang}, \citenamefont {Zeng},
		\citenamefont {Zhang}, \citenamefont {Liu}, \citenamefont {Shi},\ and\
		\citenamefont {Zi}}]{Wang2019}%
	\BibitemOpen
	\bibfield  {author} {\bibinfo {author} {\bibfnamefont {J.}~\bibnamefont
			{Wang}}, \bibinfo {author} {\bibfnamefont {A.}~\bibnamefont {Chen}}, \bibinfo
		{author} {\bibfnamefont {Y.}~\bibnamefont {Zhang}}, \bibinfo {author}
		{\bibfnamefont {J.}~\bibnamefont {Zeng}}, \bibinfo {author} {\bibfnamefont
			{Y.}~\bibnamefont {Zhang}}, \bibinfo {author} {\bibfnamefont
			{X.}~\bibnamefont {Liu}}, \bibinfo {author} {\bibfnamefont {L.}~\bibnamefont
			{Shi}}, \ and\ \bibinfo {author} {\bibfnamefont {J.}~\bibnamefont {Zi}},\
	}\href {\doibase 10.1103/physrevb.100.075407} {\bibfield  {journal} {\bibinfo
			{journal} {Phys. Rev. B}\ }\textbf {\bibinfo {volume} {100}},\ \bibinfo
		{pages} {075407} (\bibinfo {year} {2019})}\BibitemShut {NoStop}%
	\bibitem [{\citenamefont {Xiao}\ \emph {et~al.}(2020)\citenamefont {Xiao},
		\citenamefont {Liu}, \citenamefont {Wang}, \citenamefont {Liu},\ and\
		\citenamefont {Zhou}}]{Xiao2020}%
	\BibitemOpen
	\bibfield  {author} {\bibinfo {author} {\bibfnamefont {S.}~\bibnamefont
			{Xiao}}, \bibinfo {author} {\bibfnamefont {T.}~\bibnamefont {Liu}}, \bibinfo
		{author} {\bibfnamefont {X.}~\bibnamefont {Wang}}, \bibinfo {author}
		{\bibfnamefont {X.}~\bibnamefont {Liu}}, \ and\ \bibinfo {author}
		{\bibfnamefont {C.}~\bibnamefont {Zhou}},\ }\href {\doibase
		10.1103/physrevb.102.085410} {\bibfield  {journal} {\bibinfo  {journal}
			{Phys. Rev. B}\ }\textbf {\bibinfo {volume} {102}},\ \bibinfo {pages}
		{085410} (\bibinfo {year} {2020})}\BibitemShut {NoStop}%
	\bibitem [{\citenamefont {Wang}\ \emph
		{et~al.}(2020{\natexlab{a}})\citenamefont {Wang}, \citenamefont {Duan},
		\citenamefont {Chen}, \citenamefont {Zhou}, \citenamefont {Liu},\ and\
		\citenamefont {Xiao}}]{Wang2020a}%
	\BibitemOpen
	\bibfield  {author} {\bibinfo {author} {\bibfnamefont {X.}~\bibnamefont
			{Wang}}, \bibinfo {author} {\bibfnamefont {J.}~\bibnamefont {Duan}}, \bibinfo
		{author} {\bibfnamefont {W.}~\bibnamefont {Chen}}, \bibinfo {author}
		{\bibfnamefont {C.}~\bibnamefont {Zhou}}, \bibinfo {author} {\bibfnamefont
			{T.}~\bibnamefont {Liu}}, \ and\ \bibinfo {author} {\bibfnamefont
			{S.}~\bibnamefont {Xiao}},\ }\href {\doibase 10.1103/physrevb.102.155432}
	{\bibfield  {journal} {\bibinfo  {journal} {Phys. Rev. B}\ }\textbf {\bibinfo
			{volume} {102}},\ \bibinfo {pages} {155432} (\bibinfo {year}
		{2020}{\natexlab{a}})}\BibitemShut {NoStop}%
	\bibitem [{\citenamefont {Huang}\ \emph {et~al.}(2016)\citenamefont {Huang},
		\citenamefont {Li}, \citenamefont {Gurarslan}, \citenamefont {Yu},
		\citenamefont {Kirste}, \citenamefont {Guo}, \citenamefont {Zhao},
		\citenamefont {Collazo}, \citenamefont {Sitar}, \citenamefont {Parsons},
		\citenamefont {Kudenov},\ and\ \citenamefont {Cao}}]{Huang2016}%
	\BibitemOpen
	\bibfield  {author} {\bibinfo {author} {\bibfnamefont {L.}~\bibnamefont
			{Huang}}, \bibinfo {author} {\bibfnamefont {G.}~\bibnamefont {Li}}, \bibinfo
		{author} {\bibfnamefont {A.}~\bibnamefont {Gurarslan}}, \bibinfo {author}
		{\bibfnamefont {Y.}~\bibnamefont {Yu}}, \bibinfo {author} {\bibfnamefont
			{R.}~\bibnamefont {Kirste}}, \bibinfo {author} {\bibfnamefont
			{W.}~\bibnamefont {Guo}}, \bibinfo {author} {\bibfnamefont {J.}~\bibnamefont
			{Zhao}}, \bibinfo {author} {\bibfnamefont {R.}~\bibnamefont {Collazo}},
		\bibinfo {author} {\bibfnamefont {Z.}~\bibnamefont {Sitar}}, \bibinfo
		{author} {\bibfnamefont {G.~N.}\ \bibnamefont {Parsons}}, \bibinfo {author}
		{\bibfnamefont {M.}~\bibnamefont {Kudenov}}, \ and\ \bibinfo {author}
		{\bibfnamefont {L.}~\bibnamefont {Cao}},\ }\href {\doibase
		10.1021/acsnano.6b02195} {\bibfield  {journal} {\bibinfo  {journal} {{ACS}
				Nano}\ }\textbf {\bibinfo {volume} {10}},\ \bibinfo {pages} {7493} (\bibinfo
		{year} {2016})}\BibitemShut {NoStop}%
	\bibitem [{\citenamefont {Li}\ \emph {et~al.}(2017)\citenamefont {Li},
		\citenamefont {Qin}, \citenamefont {Wang}, \citenamefont {Zhai},
		\citenamefont {Ren},\ and\ \citenamefont {Hu}}]{Li2017}%
	\BibitemOpen
	\bibfield  {author} {\bibinfo {author} {\bibfnamefont {H.}~\bibnamefont
			{Li}}, \bibinfo {author} {\bibfnamefont {M.}~\bibnamefont {Qin}}, \bibinfo
		{author} {\bibfnamefont {L.}~\bibnamefont {Wang}}, \bibinfo {author}
		{\bibfnamefont {X.}~\bibnamefont {Zhai}}, \bibinfo {author} {\bibfnamefont
			{R.}~\bibnamefont {Ren}}, \ and\ \bibinfo {author} {\bibfnamefont
			{J.}~\bibnamefont {Hu}},\ }\href {\doibase 10.1364/oe.25.031612} {\bibfield
		{journal} {\bibinfo  {journal} {Opt. Express}\ }\textbf {\bibinfo {volume}
			{25}},\ \bibinfo {pages} {31612} (\bibinfo {year} {2017})}\BibitemShut
	{NoStop}%
	\bibitem [{\citenamefont {Jiang}\ \emph {et~al.}(2018)\citenamefont {Jiang},
		\citenamefont {Wang}, \citenamefont {Xiao}, \citenamefont {Yan},
		\citenamefont {Cheng},\ and\ \citenamefont {Zhong}}]{Jiang2018}%
	\BibitemOpen
	\bibfield  {author} {\bibinfo {author} {\bibfnamefont {X.}~\bibnamefont
			{Jiang}}, \bibinfo {author} {\bibfnamefont {T.}~\bibnamefont {Wang}},
		\bibinfo {author} {\bibfnamefont {S.}~\bibnamefont {Xiao}}, \bibinfo {author}
		{\bibfnamefont {X.}~\bibnamefont {Yan}}, \bibinfo {author} {\bibfnamefont
			{L.}~\bibnamefont {Cheng}}, \ and\ \bibinfo {author} {\bibfnamefont
			{Q.}~\bibnamefont {Zhong}},\ }\href {\doibase 10.1088/1361-6528/aac8f1}
	{\bibfield  {journal} {\bibinfo  {journal} {Nanotechnology}\ }\textbf
		{\bibinfo {volume} {29}},\ \bibinfo {pages} {335205} (\bibinfo {year}
		{2018})}\BibitemShut {NoStop}%
	\bibitem [{\citenamefont {Hong}\ \emph {et~al.}(2019)\citenamefont {Hong},
		\citenamefont {Chen}, \citenamefont {Zhang}, \citenamefont {Zhu},
		\citenamefont {Qin},\ and\ \citenamefont {Yuan}}]{Hong2019}%
	\BibitemOpen
	\bibfield  {author} {\bibinfo {author} {\bibfnamefont {Q.}~\bibnamefont
			{Hong}}, \bibinfo {author} {\bibfnamefont {X.}~\bibnamefont {Chen}}, \bibinfo
		{author} {\bibfnamefont {J.}~\bibnamefont {Zhang}}, \bibinfo {author}
		{\bibfnamefont {Z.}~\bibnamefont {Zhu}}, \bibinfo {author} {\bibfnamefont
			{S.}~\bibnamefont {Qin}}, \ and\ \bibinfo {author} {\bibfnamefont
			{X.}~\bibnamefont {Yuan}},\ }\href {\doibase 10.1039/c9nr06192d} {\bibfield
		{journal} {\bibinfo  {journal} {Nanoscale}\ }\textbf {\bibinfo {volume}
			{11}},\ \bibinfo {pages} {23149} (\bibinfo {year} {2019})}\BibitemShut
	{NoStop}%
	\bibitem [{\citenamefont {Cao}\ \emph {et~al.}(2020)\citenamefont {Cao},
		\citenamefont {Dong}, \citenamefont {He}, \citenamefont {Forsberg},
		\citenamefont {Jin},\ and\ \citenamefont {He}}]{Cao2020}%
	\BibitemOpen
	\bibfield  {author} {\bibinfo {author} {\bibfnamefont {S.}~\bibnamefont
			{Cao}}, \bibinfo {author} {\bibfnamefont {H.}~\bibnamefont {Dong}}, \bibinfo
		{author} {\bibfnamefont {J.}~\bibnamefont {He}}, \bibinfo {author}
		{\bibfnamefont {E.}~\bibnamefont {Forsberg}}, \bibinfo {author}
		{\bibfnamefont {Y.}~\bibnamefont {Jin}}, \ and\ \bibinfo {author}
		{\bibfnamefont {S.}~\bibnamefont {He}},\ }\href {\doibase
		10.1021/acs.jpclett.0c01080} {\bibfield  {journal} {\bibinfo  {journal} {J.
				Phys. Chem. Lett.}\ }\textbf {\bibinfo {volume} {11}},\ \bibinfo {pages}
		{4631} (\bibinfo {year} {2020})}\BibitemShut {NoStop}%
	\bibitem [{\citenamefont {Wang}\ \emph
		{et~al.}(2020{\natexlab{b}})\citenamefont {Wang}, \citenamefont {Yang},\ and\
		\citenamefont {Shi}}]{Wang2020}%
	\BibitemOpen
	\bibfield  {author} {\bibinfo {author} {\bibfnamefont {J.}~\bibnamefont
			{Wang}}, \bibinfo {author} {\bibfnamefont {J.}~\bibnamefont {Yang}}, \ and\
		\bibinfo {author} {\bibfnamefont {D.}~\bibnamefont {Shi}},\ }\href {\doibase
		10.1088/1361-6528/abaa10} {\bibfield  {journal} {\bibinfo  {journal}
			{Nanotechnology}\ }\textbf {\bibinfo {volume} {31}},\ \bibinfo {pages}
		{465205} (\bibinfo {year} {2020}{\natexlab{b}})}\BibitemShut {NoStop}%
	\bibitem [{\citenamefont {Qing}\ \emph {et~al.}(2018)\citenamefont {Qing},
		\citenamefont {Ma},\ and\ \citenamefont {Cui}}]{Qing2018}%
	\BibitemOpen
	\bibfield  {author} {\bibinfo {author} {\bibfnamefont {Y.~M.}\ \bibnamefont
			{Qing}}, \bibinfo {author} {\bibfnamefont {H.~F.}\ \bibnamefont {Ma}}, \ and\
		\bibinfo {author} {\bibfnamefont {T.~J.}\ \bibnamefont {Cui}},\ }\href
	{\doibase 10.1364/oe.26.032442} {\bibfield  {journal} {\bibinfo  {journal}
			{Opt. Express}\ }\textbf {\bibinfo {volume} {26}},\ \bibinfo {pages} {32442}
		(\bibinfo {year} {2018})}\BibitemShut {NoStop}%
	\bibitem [{\citenamefont {Xiao}\ \emph {et~al.}(2019)\citenamefont {Xiao},
		\citenamefont {Liu}, \citenamefont {Cheng}, \citenamefont {Zhou},
		\citenamefont {Jiang}, \citenamefont {Li},\ and\ \citenamefont
		{Xu}}]{Xiao2019}%
	\BibitemOpen
	\bibfield  {author} {\bibinfo {author} {\bibfnamefont {S.}~\bibnamefont
			{Xiao}}, \bibinfo {author} {\bibfnamefont {T.}~\bibnamefont {Liu}}, \bibinfo
		{author} {\bibfnamefont {L.}~\bibnamefont {Cheng}}, \bibinfo {author}
		{\bibfnamefont {C.}~\bibnamefont {Zhou}}, \bibinfo {author} {\bibfnamefont
			{X.}~\bibnamefont {Jiang}}, \bibinfo {author} {\bibfnamefont
			{Z.}~\bibnamefont {Li}}, \ and\ \bibinfo {author} {\bibfnamefont
			{C.}~\bibnamefont {Xu}},\ }\href {\doibase 10.1109/jlt.2019.2914183}
	{\bibfield  {journal} {\bibinfo  {journal} {J. Lightw. Technol.}\ }\textbf
		{\bibinfo {volume} {37}},\ \bibinfo {pages} {3290} (\bibinfo {year}
		{2019})}\BibitemShut {NoStop}%
	\bibitem [{\citenamefont {Liu}\ \emph {et~al.}(2019{\natexlab{b}})\citenamefont
		{Liu}, \citenamefont {Jiang}, \citenamefont {Zhou},\ and\ \citenamefont
		{Xiao}}]{Liu2019}%
	\BibitemOpen
	\bibfield  {author} {\bibinfo {author} {\bibfnamefont {T.}~\bibnamefont
			{Liu}}, \bibinfo {author} {\bibfnamefont {X.}~\bibnamefont {Jiang}}, \bibinfo
		{author} {\bibfnamefont {C.}~\bibnamefont {Zhou}}, \ and\ \bibinfo {author}
		{\bibfnamefont {S.}~\bibnamefont {Xiao}},\ }\href {\doibase
		10.1364/oe.27.027618} {\bibfield  {journal} {\bibinfo  {journal} {Opt.
				Express}\ }\textbf {\bibinfo {volume} {27}},\ \bibinfo {pages} {27618}
		(\bibinfo {year} {2019}{\natexlab{b}})}\BibitemShut {NoStop}%
	\bibitem [{\citenamefont {Liu}\ \emph {et~al.}(2020)\citenamefont {Liu},
		\citenamefont {Jiang}, \citenamefont {Wang}, \citenamefont {Liu},
		\citenamefont {Zhou},\ and\ \citenamefont {Xiao}}]{Liu2020}%
	\BibitemOpen
	\bibfield  {author} {\bibinfo {author} {\bibfnamefont {T.}~\bibnamefont
			{Liu}}, \bibinfo {author} {\bibfnamefont {X.}~\bibnamefont {Jiang}}, \bibinfo
		{author} {\bibfnamefont {H.}~\bibnamefont {Wang}}, \bibinfo {author}
		{\bibfnamefont {Y.}~\bibnamefont {Liu}}, \bibinfo {author} {\bibfnamefont
			{C.}~\bibnamefont {Zhou}}, \ and\ \bibinfo {author} {\bibfnamefont
			{S.}~\bibnamefont {Xiao}},\ }\href {\doibase 10.7567/1882-0786/ab6270}
	{\bibfield  {journal} {\bibinfo  {journal} {Appl. Phys. Express}\ }\textbf
		{\bibinfo {volume} {13}},\ \bibinfo {pages} {012010} (\bibinfo {year}
		{2020})}\BibitemShut {NoStop}%
	\bibitem [{\citenamefont {Xu}\ \emph {et~al.}(2020)\citenamefont {Xu},
		\citenamefont {Li}, \citenamefont {Zhang}, \citenamefont {Bai}, \citenamefont
		{Zhang},\ and\ \citenamefont {Qin}}]{Xu2020}%
	\BibitemOpen
	\bibfield  {author} {\bibinfo {author} {\bibfnamefont {Y.}~\bibnamefont
			{Xu}}, \bibinfo {author} {\bibfnamefont {H.}~\bibnamefont {Li}}, \bibinfo
		{author} {\bibfnamefont {X.}~\bibnamefont {Zhang}}, \bibinfo {author}
		{\bibfnamefont {Z.}~\bibnamefont {Bai}}, \bibinfo {author} {\bibfnamefont
			{Z.}~\bibnamefont {Zhang}}, \ and\ \bibinfo {author} {\bibfnamefont
			{S.}~\bibnamefont {Qin}},\ }\href {\doibase 10.1364/ao.405225} {\bibfield
		{journal} {\bibinfo  {journal} {Appl. Opt.}\ }\textbf {\bibinfo {volume}
			{59}},\ \bibinfo {pages} {9003} (\bibinfo {year} {2020})}\BibitemShut
	{NoStop}%
	\bibitem [{\citenamefont {Zheludev}\ \emph {et~al.}(2008)\citenamefont
		{Zheludev}, \citenamefont {Prosvirnin}, \citenamefont {Papasimakis},\ and\
		\citenamefont {Fedotov}}]{Zheludev2008}%
	\BibitemOpen
	\bibfield  {author} {\bibinfo {author} {\bibfnamefont {N.~I.}\ \bibnamefont
			{Zheludev}}, \bibinfo {author} {\bibfnamefont {S.~L.}\ \bibnamefont
			{Prosvirnin}}, \bibinfo {author} {\bibfnamefont {N.}~\bibnamefont
			{Papasimakis}}, \ and\ \bibinfo {author} {\bibfnamefont {V.~A.}\ \bibnamefont
			{Fedotov}},\ }\href {\doibase 10.1038/nphoton.2008.82} {\bibfield  {journal}
		{\bibinfo  {journal} {Nat. Photonics}\ }\textbf {\bibinfo {volume} {2}},\
		\bibinfo {pages} {351} (\bibinfo {year} {2008})}\BibitemShut {NoStop}%
	\bibitem [{\citenamefont {Xiao}\ \emph {et~al.}(2010)\citenamefont {Xiao},
		\citenamefont {Drachev}, \citenamefont {Kildishev}, \citenamefont {Ni},
		\citenamefont {Chettiar}, \citenamefont {Yuan},\ and\ \citenamefont
		{Shalaev}}]{Xiao2010}%
	\BibitemOpen
	\bibfield  {author} {\bibinfo {author} {\bibfnamefont {S.}~\bibnamefont
			{Xiao}}, \bibinfo {author} {\bibfnamefont {V.~P.}\ \bibnamefont {Drachev}},
		\bibinfo {author} {\bibfnamefont {A.~V.}\ \bibnamefont {Kildishev}}, \bibinfo
		{author} {\bibfnamefont {X.}~\bibnamefont {Ni}}, \bibinfo {author}
		{\bibfnamefont {U.~K.}\ \bibnamefont {Chettiar}}, \bibinfo {author}
		{\bibfnamefont {H.-K.}\ \bibnamefont {Yuan}}, \ and\ \bibinfo {author}
		{\bibfnamefont {V.~M.}\ \bibnamefont {Shalaev}},\ }\href {\doibase
		10.1038/nature09278} {\bibfield  {journal} {\bibinfo  {journal} {Nature}\
		}\textbf {\bibinfo {volume} {466}},\ \bibinfo {pages} {735} (\bibinfo {year}
		{2010})}\BibitemShut {NoStop}%
	\bibitem [{\citenamefont {Yoon}\ \emph {et~al.}(2015)\citenamefont {Yoon},
		\citenamefont {Jung},\ and\ \citenamefont {Song}}]{Yoon2015}%
	\BibitemOpen
	\bibfield  {author} {\bibinfo {author} {\bibfnamefont {J.~W.}\ \bibnamefont
			{Yoon}}, \bibinfo {author} {\bibfnamefont {M.~J.}\ \bibnamefont {Jung}}, \
		and\ \bibinfo {author} {\bibfnamefont {S.~H.}\ \bibnamefont {Song}},\ }\href
	{\doibase 10.1364/ol.40.002309} {\bibfield  {journal} {\bibinfo  {journal}
			{Opt. Lett.}\ }\textbf {\bibinfo {volume} {40}},\ \bibinfo {pages} {2309}
		(\bibinfo {year} {2015})}\BibitemShut {NoStop}%
	\bibitem [{\citenamefont {Vasi{\'{c}}}\ and\ \citenamefont
		{Gaji{\'{c}}}(2017)}]{Vasic2017}%
	\BibitemOpen
	\bibfield  {author} {\bibinfo {author} {\bibfnamefont {B.}~\bibnamefont
			{Vasi{\'{c}}}}\ and\ \bibinfo {author} {\bibfnamefont {R.}~\bibnamefont
			{Gaji{\'{c}}}},\ }\href {\doibase 10.1364/ol.42.002181} {\bibfield  {journal}
		{\bibinfo  {journal} {Opt. Lett.}\ }\textbf {\bibinfo {volume} {42}},\
		\bibinfo {pages} {2181} (\bibinfo {year} {2017})}\BibitemShut {NoStop}%
	\bibitem [{\citenamefont {Wang}\ \emph {et~al.}(2017)\citenamefont {Wang},
		\citenamefont {Han}, \citenamefont {Chen}, \citenamefont {Dai}, \citenamefont
		{Zhou}, \citenamefont {Hu}, \citenamefont {Yu}, \citenamefont {Liu},
		\citenamefont {Shi},\ and\ \citenamefont {Zi}}]{Wang2017}%
	\BibitemOpen
	\bibfield  {author} {\bibinfo {author} {\bibfnamefont {J.}~\bibnamefont
			{Wang}}, \bibinfo {author} {\bibfnamefont {D.}~\bibnamefont {Han}}, \bibinfo
		{author} {\bibfnamefont {A.}~\bibnamefont {Chen}}, \bibinfo {author}
		{\bibfnamefont {Y.}~\bibnamefont {Dai}}, \bibinfo {author} {\bibfnamefont
			{M.}~\bibnamefont {Zhou}}, \bibinfo {author} {\bibfnamefont {X.}~\bibnamefont
			{Hu}}, \bibinfo {author} {\bibfnamefont {Z.}~\bibnamefont {Yu}}, \bibinfo
		{author} {\bibfnamefont {X.}~\bibnamefont {Liu}}, \bibinfo {author}
		{\bibfnamefont {L.}~\bibnamefont {Shi}}, \ and\ \bibinfo {author}
		{\bibfnamefont {J.}~\bibnamefont {Zi}},\ }\href {\doibase
		10.1103/physrevb.96.195419} {\bibfield  {journal} {\bibinfo  {journal} {Phys.
				Rev. B}\ }\textbf {\bibinfo {volume} {96}},\ \bibinfo {pages} {195419}
		(\bibinfo {year} {2017})}\BibitemShut {NoStop}%
	\bibitem [{\citenamefont {Meng}\ \emph {et~al.}(2019)\citenamefont {Meng},
		\citenamefont {Zhao}, \citenamefont {Yang}, \citenamefont {de~Abajo},
		\citenamefont {Li}, \citenamefont {Ruan},\ and\ \citenamefont
		{Qiu}}]{Meng2019}%
	\BibitemOpen
	\bibfield  {author} {\bibinfo {author} {\bibfnamefont {L.}~\bibnamefont
			{Meng}}, \bibinfo {author} {\bibfnamefont {D.}~\bibnamefont {Zhao}}, \bibinfo
		{author} {\bibfnamefont {Y.}~\bibnamefont {Yang}}, \bibinfo {author}
		{\bibfnamefont {F.~J.~G.}\ \bibnamefont {de~Abajo}}, \bibinfo {author}
		{\bibfnamefont {Q.}~\bibnamefont {Li}}, \bibinfo {author} {\bibfnamefont
			{Z.}~\bibnamefont {Ruan}}, \ and\ \bibinfo {author} {\bibfnamefont
			{M.}~\bibnamefont {Qiu}},\ }\href {\doibase 10.1103/physrevapplied.11.044030}
	{\bibfield  {journal} {\bibinfo  {journal} {Phys. Rev. Appl}\ }\textbf
		{\bibinfo {volume} {11}},\ \bibinfo {pages} {044030} (\bibinfo {year}
		{2019})}\BibitemShut {NoStop}%
	\bibitem [{\citenamefont {Sanders}\ and\ \citenamefont
		{Manjavacas}(2020)}]{Sanders2020}%
	\BibitemOpen
	\bibfield  {author} {\bibinfo {author} {\bibfnamefont {S.}~\bibnamefont
			{Sanders}}\ and\ \bibinfo {author} {\bibfnamefont {A.}~\bibnamefont
			{Manjavacas}},\ }\href {\doibase 10.1515/nanoph-2019-0392} {\bibfield
		{journal} {\bibinfo  {journal} {Nanophotonics}\ }\textbf {\bibinfo {volume}
			{9}},\ \bibinfo {pages} {473} (\bibinfo {year} {2020})}\BibitemShut {NoStop}%
	\bibitem [{\citenamefont {Palik}(1998)}]{Palik1998}%
	\BibitemOpen
	\bibfield  {author} {\bibinfo {author} {\bibfnamefont {E.~D.}\ \bibnamefont
			{Palik}},\ }\href@noop {} {\emph {\bibinfo {title} {Handbook of optical
				constants of solids}}}\ (\bibinfo  {publisher} {Academic press},\ \bibinfo
	{year} {1998})\BibitemShut {NoStop}%
	\bibitem [{\citenamefont {Stauber}\ \emph {et~al.}(2008)\citenamefont
		{Stauber}, \citenamefont {Peres},\ and\ \citenamefont {Geim}}]{Stauber2008}%
	\BibitemOpen
	\bibfield  {author} {\bibinfo {author} {\bibfnamefont {T.}~\bibnamefont
			{Stauber}}, \bibinfo {author} {\bibfnamefont {N.~M.~R.}\ \bibnamefont
			{Peres}}, \ and\ \bibinfo {author} {\bibfnamefont {A.~K.}\ \bibnamefont
			{Geim}},\ }\href {\doibase 10.1103/physrevb.78.085432} {\bibfield  {journal}
		{\bibinfo  {journal} {Phys. Rev. B}\ }\textbf {\bibinfo {volume} {78}},\
		\bibinfo {pages} {085432} (\bibinfo {year} {2008})}\BibitemShut {NoStop}%
	\bibitem [{\citenamefont {Hanson}(2008)}]{Hanson2008}%
	\BibitemOpen
	\bibfield  {author} {\bibinfo {author} {\bibfnamefont {G.~W.}\ \bibnamefont
			{Hanson}},\ }\href {\doibase 10.1063/1.2891452} {\bibfield  {journal}
		{\bibinfo  {journal} {J. Appl. Phys.}\ }\textbf {\bibinfo {volume} {103}},\
		\bibinfo {pages} {064302} (\bibinfo {year} {2008})}\BibitemShut {NoStop}%
	\bibitem [{\citenamefont {Oughstun}\ and\ \citenamefont
		{Cartwright}(2003)}]{Oughstun2003}%
	\BibitemOpen
	\bibfield  {author} {\bibinfo {author} {\bibfnamefont {K.}~\bibnamefont
			{Oughstun}}\ and\ \bibinfo {author} {\bibfnamefont {N.}~\bibnamefont
			{Cartwright}},\ }\href {\doibase 10.1364/oe.11.001541} {\bibfield  {journal}
		{\bibinfo  {journal} {Opt. Express}\ }\textbf {\bibinfo {volume} {11}},\
		\bibinfo {pages} {1541} (\bibinfo {year} {2003})}\BibitemShut {NoStop}%
	\bibitem [{\citenamefont {Miroshnichenko}\ \emph {et~al.}(2010)\citenamefont
		{Miroshnichenko}, \citenamefont {Flach},\ and\ \citenamefont
		{Kivshar}}]{Miroshnichenko2010}%
	\BibitemOpen
	\bibfield  {author} {\bibinfo {author} {\bibfnamefont {A.~E.}\ \bibnamefont
			{Miroshnichenko}}, \bibinfo {author} {\bibfnamefont {S.}~\bibnamefont
			{Flach}}, \ and\ \bibinfo {author} {\bibfnamefont {Y.~S.}\ \bibnamefont
			{Kivshar}},\ }\href {\doibase 10.1103/revmodphys.82.2257} {\bibfield
		{journal} {\bibinfo  {journal} {Rev. Mod. Phys.}\ }\textbf {\bibinfo {volume}
			{82}},\ \bibinfo {pages} {2257} (\bibinfo {year} {2010})}\BibitemShut
	{NoStop}%
	\bibitem [{\citenamefont {Zhou}\ \emph {et~al.}(2019)\citenamefont {Zhou},
		\citenamefont {Li}, \citenamefont {Wang},\ and\ \citenamefont
		{Zhan}}]{Zhou2019}%
	\BibitemOpen
	\bibfield  {author} {\bibinfo {author} {\bibfnamefont {C.}~\bibnamefont
			{Zhou}}, \bibinfo {author} {\bibfnamefont {S.}~\bibnamefont {Li}}, \bibinfo
		{author} {\bibfnamefont {Y.}~\bibnamefont {Wang}}, \ and\ \bibinfo {author}
		{\bibfnamefont {M.}~\bibnamefont {Zhan}},\ }\href {\doibase
		10.1103/physrevb.100.195306} {\bibfield  {journal} {\bibinfo  {journal}
			{Phys. Rev. B}\ }\textbf {\bibinfo {volume} {100}},\ \bibinfo {pages}
		{195306} (\bibinfo {year} {2019})}\BibitemShut {NoStop}%
	\bibitem [{\citenamefont {Li}\ and\ \citenamefont {Xia}(2010)}]{Li2010}%
	\BibitemOpen
	\bibfield  {author} {\bibinfo {author} {\bibfnamefont {Z.-Y.}\ \bibnamefont
			{Li}}\ and\ \bibinfo {author} {\bibfnamefont {Y.}~\bibnamefont {Xia}},\
	}\href {\doibase 10.1021/nl903409x} {\bibfield  {journal} {\bibinfo
			{journal} {Nano Lett.}\ }\textbf {\bibinfo {volume} {10}},\ \bibinfo {pages}
		{243} (\bibinfo {year} {2010})}\BibitemShut {NoStop}%
	\bibitem [{\citenamefont {Liu}\ \emph {et~al.}(2011)\citenamefont {Liu},
		\citenamefont {Li}, \citenamefont {Zhou}, \citenamefont {Gan},\ and\
		\citenamefont {Li}}]{Liu2011}%
	\BibitemOpen
	\bibfield  {author} {\bibinfo {author} {\bibfnamefont {S.-Y.}\ \bibnamefont
			{Liu}}, \bibinfo {author} {\bibfnamefont {J.}~\bibnamefont {Li}}, \bibinfo
		{author} {\bibfnamefont {F.}~\bibnamefont {Zhou}}, \bibinfo {author}
		{\bibfnamefont {L.}~\bibnamefont {Gan}}, \ and\ \bibinfo {author}
		{\bibfnamefont {Z.-Y.}\ \bibnamefont {Li}},\ }\href {\doibase
		10.1364/ol.36.001296} {\bibfield  {journal} {\bibinfo  {journal} {Opt.
				Lett.}\ }\textbf {\bibinfo {volume} {36}},\ \bibinfo {pages} {1296} (\bibinfo
		{year} {2011})}\BibitemShut {NoStop}%
\end{thebibliography}

%

\end{document}